\documentclass[preprint,10pt]{elsarticle}

\usepackage[english]{babel}
\usepackage{graphicx}
\usepackage[T1]{fontenc}
\usepackage{float}
\usepackage{amsfonts}
\usepackage{amsmath}
\usepackage{fancyhdr}
\usepackage{enumerate}
\usepackage{algorithm_OF} 
\usepackage{algpseudocode}

\def\s{\Sigma}
\def\b{\beta}

\begin{document}

\begin{frontmatter}
\title{Estimation of covariance matrices based on hierarchical inverse-Wishart priors\tnoteref{t1}}
\tnotetext[t1]{This document is a part of the thesis of Mathilde Bouriga.}

\author[fn1,fn2]{M. Bouriga\corref{cor1}}
\ead{mathilde.bouriga@edf.fr}
\author[fn1,fn2]{O. F\'eron}
\ead{olivier-2.feron@edf.fr}

\address[fn1]{\textit{Universit\'e Paris IX - Dauphine, Place du Maréchal de Lattre de Tassigny, 75116 Paris, France}}
\address[fn2]{\textit{EDF R\&D, 1 avenue Général de Gaulle, 92140 Clamart, France}}
\cortext[cor1]{Corresponding author. Fax: 0147653900.}

\begin{abstract}
This paper focuses on Bayesian shrinkage for covariance matrix estimation. We examine posterior properties and frequentist risks of Bayesian estimators based on new hierarchical inverse-Wishart priors.  More precisely, we give the existence conditions of the posterior distributions. 
Advantages in terms of numerical simulations of posteriors are shown. A simulation study illustrates the performance of the estimation procedures under three loss functions for relevant sample sizes and various covariance structures.
\end{abstract}


\end{frontmatter}

\section{Introduction}

\indent Estimating a covariance matrix efficiently is an important statistical issue. Often, applied scientific problems require an estimate of a covariance matrix in the context of a large matrix dimension $p$ relative to the number of observations $n$. In such settings, standard estimators - the sample covariance matrix or the maximum likelihood estimator - are known to perform poorly \cite{Stein56, Stein75, Dem69}. 
When $n$ is smaller than $p$, they are not positive definite. When it is larger, they are invertible but still inappropriate because unstable, 
unless $\frac{p}{n}$ is negligibly small. Indeed, if $n$ is of the same order as $p$, the sample eigenvalues significantly deviate from to the population eigenvalues \cite{Stein75,Dem69}. This fact has incited many authors to focus on the eigenvalues attempting to overcome their distortion. \\
Some approaches to more stably estimating the matrix in small samples have been proposed, such as shrinkage methods on which we will focus. 
The work along these directions can be found in both frequentist and Bayesian frameworks. The proposed estimators for the covariance matrixare then derived from a decision-theoretic perspective or associated with an appropriate prior and a specific loss function.\\
James and Stein \cite{James61} were the first to propose biased estimators for covariance matrixunder Stein's loss function, dominating the classical sample covariance matrix. Since, many authors have explored improved James-Stein type estimators under Stein's loss \cite{Stein75,Dey85,Kub99,Lin84} or other losses \cite{Efron74,Haff79,Kub04,Lin84}. Ledoit and Wolf \cite{Ledoit02} consider Steinian shrinkage toward the single-index covariance matrix to estimate the covariance matrix of stock returns. Furrer and Bengtsson \cite{Furrer07} consider "tapering" the sample covariance matrix, that is, gradually shrinking the off-diagonal elements toward zero.  

From a Bayesian perspective, the common approach \cite{Haff80,Champion01} yielding estimators which shrink towards a structure uses the conjugate prior on the covariance matrix, an inverse-Wishart distribution with some degrees of freedom and a scale matrix as hyperparameters. The appeal of conjugate priors is to allow efficient posterior simulations but such priors might be contested because of their lack of flexibility; specifically, with an inverse-Wishart prior, only one parameter does control the variability of the matrix elements. Efficient Bayesian estimators involving more flexible priors are obtained using various decompositions, the most well-known are derived from the variance-correlation strategy introduced by Barnard and al. \cite{Bar00,Daniels99}, the spectral \cite{Daniels99,Berger94} or the Cholesky \cite{Kohn02} decompositions of the covariance matrix and the matrix-logarithmic covariance model \cite{Hsu92}. 
In particular, in \cite{Bar00}, the covariance matrixis modeled in terms of standard deviations and correlations. After discussion about suitable priors, the authors advocate a flat prior on the space of correlation matrices and log normal priors on the variances. 
Yang and Berger \cite{Berger94} develop a reference prior approach for the covariance matrix, approach known to outperform another common noninformative prior, Jeffreys prior \cite{Jef61}, for high-dimensional problems.  
Working with the spectral decomposition of the matrix, their method shrinks the eigenvalues, it results in a better estimation of the underlying eigenstructure.
Smith and Kohn \cite{Kohn02} use a prior that allows for zero elements in the strict lower triangle of the Cholesky decomposition of the inverse of the covariance matrix. Leonard and Hsu \cite{Hsu92} model the matrix logarithm of the covariance matrix and place a multivariate normal distribution on the $\frac{p(p+1)}{2}$ vectorized non-redundant elements of this matrix.
Daniels and Kass \cite{Daniels99} focus on shrinking the matrix toward a structure, specifically, a diagonal matrix, using a fully Bayesian approach and using three different priors: normal priors on the z-transform of the correlations, normal priors on the logit of the Givens angles and inverse-Wishart prior. These parametrizations do not have simple statistical interpretation and the use of these methods has been limited in view of the difficulties of computation implied. \\
Moreover all these parametrizations do not solve the difficulty in handling and estimating hyperparameters. Empirical Bayes estimates of the hyperparameters can be proposed, for instance by \cite{Haff80,Champion01,Scha05} in the case of inverse-Wishart prior. An alternative to this empirical specification is to use hierarchical modeling. Multilevel modeling will ensure that uncertainty in higher level parameters propagates into inferences on lower level parameters and is supposed to be more flexible and also more stable than those based on diffuse priors \cite{Gelman06}. Applied to the case of inverse-Wishart prior, it is intended to offer objectivity in terms of how close the true matrix is to the specified structure by allowing for data-dependent shrinkage towards this structure. To our knowledge, hierarchical modelization based on inverse-Wishart priors has only been proposed in \cite{Daniels99} with a constraint of fixing an upper bound on the degrees of freedom to ensure proper posterior distributions. \\
In this paper we focus on hierarchical inverse-Wishart priors on the covariance matrix such that shrinkage toward diagonality is involved. This approach builds on the works of Daniels and Kass \cite{Daniels99}. After relaxing the prior on the degrees of freedom, we establish the precise conditions to ensure the properness of the posterior distributions. We also give a detailed experimental comparison of the competing Bayesian models and the maximum likelihood estimator under different loss functions. According to the loss, the priors don't have the same effect on posterior inference and we outline the limits of the hierarchical inverse-Wishart priors as a "default" choice for the covariance matrix estimation. 
\\  
The paper is structured as follows. Section \ref{sec_model} presents the Normal inverse-Wishart model as covariance shrinkage modeling approach. 
In Section \ref{sec_prior} we define the three hierarchical priors that we will consider and conditions to get proper posterior distributions. Section \ref{sec_computation} describes the attractive Markov Chain Monte Carlo sampling scheme for Bayesian computation of posteriors. 
Section \ref{sec_simulation} reports numerical results for the matrix estimators and for the eigenvalues estimators. The last section %
presents some conclusions and provides recommendations for using such priors.

\section{The normal inverse-Wishart model}

\label{sec_model}
\indent Let \textbf{X} = ($X^{(1)}$, ..., $X^{(p)}$)$^T$ be a $p$-dimensional random vector following a multivariate normal distribution $\mathcal{N}_p$(0,$\s$). 
$\s$ is an unknown covariance matrix and belongs to the set of $p \times p$ symmetric positive definite matrices $\mathcal{S}^+$. Given an independent and identically distributed sample ($\textbf{X}_1,. . . ,\textbf{X}_n$) of \textbf{X}, we wish to estimate the covariance matrix. The associated likelihood function for $\s$ is
\begin{equation} 
L(\s | S) = \frac{|\s|^{-\frac{n}{2}}}{(2 \pi)^{\frac{np}{2}}} \exp \left\{- \frac{1}{2} tr \left(\s^{-1} S\right)\right\} \label{likelihood}
\end{equation}
where $S = \sum^{n}_{i=1} \textbf{X}_i^T \textbf{X}_i$ is the scatter matrix. \\
\noindent The maximum likelihood estimator of $\s$, $\hat{\s}_{MLE}$, defined by $\frac{S}{n}$, is a classical estimator of $\s$. 
However, it becomes unstable when $p$ is moderate or large relative to the sample size $n$ because of the large number - $\frac{p(p + 1)}{2}$ - of unknown parameters to be estimated. When $n$ < $p$, $\hat{\s} _{MLE}$ is no longer positive definite. Even when $n$ > $p$, the matrix $S$ is positive definite but does lead to the distortion of the eigenstructure \cite{Stein75,Dem69} for $p$ close to $n$, especially when the true matrix is close to be diagonal. That motivates the choice of a Bayesian regularization approach and leads to assign a prior on $\s$.\\
In the absence of reliable prior information, the selection of the prior distribution is quite delicate and generic solutions must be chosen instead. Here we consider a general class of priors. 
Since the model \textbf{X} when $\s^{-1}$ varies is from a natural exponential family of distributions, a commonly used class of distributions for the canonical parameter $\s^{-1}$ is the conjugate family as defined by Diaconis \& Ylvisaker \cite{Diac79}, called the Wishart distribution. The induced prior on $\s$ 
is then the inverse-Wishart.\\
In the notation of Eaton \cite{Eat83}, $\s |\b, D \sim \mathcal{IW}(\b, D)$ means that $\s$ has the inverse-Wishart distribution with  degrees of freedom $\b > p-1$ and scale matrix $D \in \mathcal{S}^+$. The density is given by 
\begin{equation} 
\pi(\s | \b, D) = \frac{D^{\frac{\b}{2}}}{C(\b, p)}|\s|^{-\frac{\b +p +1}{2}} \exp \left\{- \frac{1}{2} tr \left(\s^{-1} D\right)\right\}. \label{inv_wishart}
\end{equation}
The normalising constant turns out to be equal to 
\begin{equation} 
C(\b, p) = 2^{\frac{\b p}{2}} \Gamma_p\left(\frac{\b}{2}\right)
\end{equation}
where $\Gamma_p(.)$ is the multivariate Gamma function defined as $\Gamma_p(a) = \pi^{\frac{p(p-1)}{4}} \prod^{p}_{j=1}\Gamma\left(a+\frac{1-j}{2}\right)$. The restriction that $\b$ be greater than $p-1$ is necessary so that $\Gamma\left(\frac{\b+1-j}{2}\right)$ be well defined. Moreover, as $\mathbb{E}(\s)$= $\frac{D}{\b-p-1}$, the expectation of $\s$ will exist if and only if $\b>p+1$.\\
\\
Now, let ($\textbf{X}_1,. . . ,\textbf{X}_n$) be a Gaussian sample associated with an inverse-Wishart prior on $\s$ centered in $D$, 
the posterior mean of $\s$, a likely estimator of $\s$, is then equal to
\begin{center}
$\mathbb{E}(\s|S)$ = $\frac{(\b-p-1) D + S}{\b + n - p - 1}$ = $\frac{(\b-p-1) D + n \hat{\Sigma}_{MLE}}{\b + n - p -1}$ \ \text{with $\b >p+1$}.\\
\end{center}
It shows that $\beta$ controls the amount of shrinkage: when $n$ is held fixed and $\b$ is allowed to grow, the posterior mean tends towards $D$ while the estimator tends towards $\hat{\Sigma}_{MLE}$ if $\b$ is held fixed and $n$ is allowed to grow. \\
When we consider the eigenvalues of the posterior mean, it is easy to see that the eigenvalues $g_i$, $i$ = 1,..., $p$ of $\mathbb{E}(\s|S)$ are 
\begin{equation}
g_i=\frac{(\b-p-1)d_{ii} + n l_i}{\b + n - p -1}, \ \forall i=1,...,p.
\end{equation}
where $l_i$ are the eigenvalues of $\hat{\Sigma}_{MLE}$.\\
We can check that, for $l_i < d_{ii}$, we always have $l_i < g_i$ and, for $l_i > 1$, we have $l_i > g_i$. \\
Notice that we can consider $\mathbb{E}(\s^{-1}|S)^{-1}$ as the estimator of $\s$, you will find the same kind of results.\\
In summary, the span of the eigenvalues of Bayes estimators based on inverse-Wishart prior will be smaller than the span of the eigenvalues of $S$, which could be used to correct the instability of the standard estimators.

\section{Inverse-Wishart prior distribution: choice of the hyperparameters}
\label{sec_prior}
\indent As seen in Section \ref{sec_model} the inverse-Wishart distribution is characterized by the hyperparameters $\b$ and $D$. 
Sometimes the hyperparameters can be specified by the investigators but there is rarely good scientific information on which to base these specifications. 
They can be obtained by empirical Bayes estimation \cite{Haff80,Champion01,Scha05}. A frequently-applied procedure is to set the scale matrix $D$ equal to $\hat{\Sigma}_{MLE}$ and $p$ degrees of freedom. 
However it turns out to be not convenient as soon as $\hat{\Sigma}_{MLE}$ becomes suspect.\\
Consequently, we prefer to embed a structure for $D$. The most commonly employed matrix targets are the identity matrix and its scalar multiples \cite{Scha05}. They are low-dimensional, thus they impose a rather strong structure which in turn requires only little data to fit the hyperparameters remaining to estimate. Alternatively we can assign a further prior distribution on these hyperparameters. This implies hierarchical models which allow a more objective approach to inference \cite{Gelman06}. They are supposed to provide more flexibility than non-hierarchical priors. As for the degrees of freedom $\b$, they can either be taken as small as possible with the idea that large values supporting the scale matrix structure or they can be given their own prior distribution.\\
\\
Here we adopt a fully Bayesian approach: we investigate three inverse-Wishart hierarchical priors with unknown degrees of freedom and scale matrix, inducing shrinkage towards diagonality.

\subsection{Daniels and Kass prior}
\label{sec_model_DK}
Daniels and Kass \cite{Daniels99} assign flat improper priors on the logarithm of the elements of the diagonal scale matrix and a vague proper uniform distribution on the logarithm of the degrees of freedom, over $]p-1 ; b ]$, with $b$ a large value. \\
Let $A$ be a $p\times p$ diagonal definite-positive matrix, the Daniels and Kass model, namely Model D\&K, is defined by
\begin{equation}
\begin{split}
& \Sigma |\alpha_j,\beta  \sim  \mathcal{IW}(\beta,A), \ A=diag(\alpha_1,...,\alpha_p)\\
& \pi(\alpha_j)  \propto  \frac{1}{\alpha_j}\mathbb{I}_{]0;+\infty[} (\alpha_j)\\
& \pi(\beta) \propto \frac{1}{\beta}\mathbb{I}_{]p-1;b]} (\beta).
\label{DK}
\end{split} 
\end{equation}

\noindent From (\ref{likelihood}) and (\ref{DK}), the joint posterior distribution given the data $\textbf{X}$ is equal to
\begin{equation}
\pi(\Sigma,A,\beta | S) = \frac{\left|\Sigma^{-1}\right|^{\frac{\beta+n+p+1}{2}} \exp\left[-\frac{1}{2}tr(\Sigma^{-1}(S+A))\right] |A|^{\frac{\beta}{2}-1}}{(2\pi)^{\frac{pn}{2}}2^{\frac{p\beta}{2}} \beta \Gamma_p(\frac{\beta}{2})}\mathbb{I}_{\textsl{S}^+ \times (\mathbb{R^*_+})^p \times]p-1;b[} (\s, \alpha_1,...,\alpha_p, \beta). \label{eq:aposterioriJointe0}
\end{equation}

\noindent The bound $b$ must be a finite value to keep the joint posterior distribution proper, with constraint to be large in order to minimize its effect on inference. Anyway we exclude arbitrarily some values of $\b$: we never know how large is large enough. Another similar but improper prior on the logarithm of $\b$ is preferred, rather than a proper one with bounded support.

\subsection{A "diagonal, equal variance" model as prior mean matrix}
\label{sec_model_1}
We choose to center the prior matrix in $\alpha I_p$. By this way, the resulting estimators will shrink all elements of $S$. Model 1 will be defined by

\begin{equation}
\begin{split}
& \Sigma |\alpha,\beta \sim \mathcal{IW}(\beta, (\beta - p -1)\alpha I_p)\\
& \pi(\alpha) \propto \frac{1}{\alpha}\mathbb{I}_{]0;+\infty[} (\alpha)\\
& \pi(\beta) \propto \frac{1}{\beta^{\delta}}\mathbb{I}_{]p+1;+\infty[} (\beta) \ \ \text{, with} \ \delta \ \text{positive integer.}
\label{eq:model1}
\end{split} 
\end{equation}

\noindent After a reparametrization given by $\phi=\alpha(\beta-p-1)$, we derive (\ref{eq:aposterioriJointe3_bis}), the joint posterior distribution of $(\Sigma, \phi, \beta)$ given $ S$:
\begin{equation}
\pi(\Sigma,\phi,\beta | S) = \frac{\left|\Sigma^{-1}\right|^{\frac{\beta+n+p+1}{2}} \exp\left[-\frac{1}{2}tr(\Sigma^{-1}(S+\phi I))\right] \left|\phi I\right|^{\frac{\beta }{2}}}{(2\pi)^{\frac{pn}{2}}2^{\frac{p\beta}{2}} \beta^{\delta} \Gamma_p(\frac{\beta}{2})\phi}\mathbb{I}_{\textsl{S}^+ \times \mathbb{R^*_+} \times]p+1;+\infty[} (\s, \phi, \beta).
\label{eq:aposterioriJointe3_bis}
\end{equation}
\noindent The introduction of $\delta$ permits the limit superior of $\beta$ in Model D\&K to be relaxed. The problem of parameter choice persists but it is hoped that this model influences less the posterior distribution. \\
The conditions ensuring properness of the posterior distribution are driven by $\delta$. For $\delta=1$, we know by \cite{Daniels99} that (\ref{eq:aposterioriJointe3_bis}) is improper.\\

\noindent \textbf{Proposition 1}: \textit{The posterior distribution (\ref{eq:aposterioriJointe3_bis}) is a proper probability density function for all} $\delta > 1$:
\begin{equation}
\int^{+\infty}_{p+1} \int_{]0;+\infty[} \int_{\textsl{S}^+} \pi(\Sigma,\phi,\beta  | S) d\Sigma d\phi d\beta  < \infty
\end{equation}\label{eq:condition_proper}

\noindent 
This proof is deferred to Appendix \ref{annexe_loi_propre}.

\noindent It leads to a remark about the posterior distribution of $\beta$.\\
\\
\textbf{Corollary}:\textit{ The posterior marginal distribution of $\beta$, $\pi(\beta|S)$, has its m-th moment and higher ones defined if $m < \delta-1$}. \\
\\
Consequently, if $\beta$ is no longer considered as a nuisance parameter but as a parameter of interest, samples from $\pi(\beta| \textbf{X})$ must be used with caution. 
For instance, for $\delta < 3$, the mean of sample paths from $\pi(\beta| \textbf{X})$ has no sense since the posterior mean of $\beta$ will not exist. 

\subsection{A "diagonal, unequal variance" model as prior mean matrix}
\label{sec_model_2}
We can prefer to keep a less strong structure for the matrix target, as in Section \ref{sec_model_DK}, where we only shrink the off-diagonal elements of $S$. Moreover we assume that the prior mean matrix does exists, equal to an arbitrary diagonal matrix $A$. Let $A$ be a $p\times p$ diagonal definite-positive matrix, Model 2 will be defined by
\begin{equation}
\begin{split}
& \Sigma |\alpha_j,\beta \sim \mathcal{IW}(\beta, (\beta - p -1)A), \ A=diag(\alpha_1,...,\alpha_p) \\ 
& \pi(\alpha_j) \propto \frac{1}{\alpha_j}\mathbb{I}_{]0;+\infty[} (\alpha_j)\\
& \pi(\beta) \propto \frac{1}{\beta^{\delta}}\mathbb{I}_{]p+1;+\infty[} (\beta).
\end{split} \label{model2}
\end{equation}

\noindent Proceed to the change of variable $\Phi=(\beta-p-1)A$ and refer the diagonal matrix $\Phi$ by its diagonal elements, $\phi_i $, $i$ from $1$ to $p$. The joint posterior distribution is then given by
\begin{equation}
\pi(\Sigma,\Phi,\beta | S) = \frac{\left|\Sigma^{-1}\right|^{\frac{\beta+n+p+1}{2}} \exp\left[-\frac{1}{2}tr(\Sigma^{-1}(S+\Phi ))\right] \left|\Phi\right|^{\frac{\beta }{2}-1}}{(2\pi)^{\frac{pn}{2}}2^{\frac{p\beta}{2}} \beta^{\delta} \Gamma_p(\frac{\beta}{2})}\mathbb{I}_{\textsl{S}^+ \times (\mathbb{R^*_+})^p \times]p+1;+\infty[}(\s, \phi_1,...,\phi_p, \beta).
\label{eq:aposterioriJointe3_mod2}
\end{equation}
\textbf{Proposition 2}: \textit{The posterior distribution (\ref{eq:aposterioriJointe3_mod2}) is a proper probability density function for all} $\delta > 1$:
\begin{equation}
\int^{+\infty}_{p+1} \int_{]0;+\infty[^p} \int_{\textsl{S}^+} \pi(\Sigma,\Phi,\beta  | S) d\Sigma d\phi_1...d\phi_p d\beta  < \infty
\end{equation}\label{eq:condition_proper_bis}

\noindent The proof of Proposition 2 is deferred to the Appendix \ref{annexe_loi_propre}.
Note that the posterior distribution (\ref{eq:aposterioriJointe3_mod2}) for Model 2 appears quasi-identical to the density (\ref{eq:aposterioriJointe0}) for Model D\&K, although the prior hypotheses differ. The existence of the prior mean matrix is only taken into account in the lower bound on $\b$ in the posterior distribution. The second difference is in the power of $\b$ in the denominator.

\section{Computation of Bayes estimators}
\label{sec_computation}
\indent Now we can derive the Bayes estimators related to the prior models of Sections \ref{sec_model_1} and \ref{sec_model_2}, under two loss functions. We will use Markov Chain Monte Carlo (MCMC) simulations to estimate these posterior quantities numerically.

\subsection{Loss functions and associated estimators}
\label{subsec_loss}
A question that naturally arises in various contexts in multivariate analysis and related topics is whether to estimate $\s$ or its inverse. We choose to focus on the estimation of $\s$ in this paper. This parameter has a natural and well understood interpretation in multivariate analysis and its direct estimation has numerous applications.\\
In the following, the Bayes estimators of $\s$ are calculated with respect to two common loss functions:
\begin{itemize}
	\item the squared Frobenius loss function
\begin{equation*}
L_2(\hat{\s},\s) =\text{tr}(\hat{\s}-\s)^2
\end{equation*}
	\item Stein's loss function 
\begin{equation*}
L_1(\hat{\s},\s)=\text{tr}(\hat{\s}\s ^{-1})- \text{log} \det (\hat{\s}\s ^{-1})-p
\end{equation*}
\end{itemize}
The corresponding Bayes estimators for $\s$ are, respectively, $\hat{\Sigma}_2 = \mathbb{E}\left(\s | S\right)$ and $\hat{\Sigma}_1 = \mathbb{E}\left(\s^{-1} | S\right)^{-1}$.\\
\\
The $L_2$ loss corresponds to the equivalent of the squared error loss function in a matrix setting. Thus $L_2(\hat{\s},\s)=\sum^{p}_{i=1}\sum^{p}_{i=1} (\hat{\sigma}_{ij} -\sigma_{ij})^2$ is a natural quadratic measure of distance between the true ($\s$) and inferred covariance matrix ($\hat{\s}$). 
\\
The $L_1$ loss was introduced by Stein \cite{Stein75} to estimate the multinormal covariance matrix and also called entropy loss. It results from evaluating the divergence of Kullback-Leibler, namely $\int p(x) log\left\{\frac{p(x)}{q(x)}\right\}dx$ for two Gaussian distributions with densities $p(x)$ and $q(x)$ defined by covariance matrices $\Sigma$ and $\hat{\Sigma}$. This scale invariant loss function will penalize the relative estimation error of the small eigenvalues, as illustrated in Section \ref{sec_simulation}. Various alternative losses have also been proposed in the literature \cite{Berger94}.\\
Both estimators are approximated by using the MCMC sampling algorithm described in Section \ref{sec_MCMC}.

\subsection{MCMC algorithm for sampling posterior distributions}
\label{sec_MCMC}
Once we introduce hierarchical priors, the conjugate structure that typically makes Gibbs sampling so attractive is no longer ensured. Fortunately, since the inverse-Wishart distributions is also a conditionally-conjugate family, the full conditional distribution of $\s$ is still inverse-Wishart. Moreover the full conditional for the additional parameters $\Phi$ and $\b$ can be also simulated easily.
\\
Sampling from the target posterior will call on iterative resampling from inverse-Wishart, inverse-Gamma distributions and from the posterior distribution of $\beta$. The simulation of the latter is based on a Metropolis sampling scheme, described in Algorithm \ref{alg:algo_mod3}.\\
\\
A finite-sample distribution from the posterior distribution (\ref{eq:aposterioriJointe3_bis}) of $(\Sigma,\phi,\beta)$ from Model 1 can then be obtained by a systematic scan Metropolis-Hasting-within-Gibbs algorithm \cite{Rob06}, described by Algorithm \ref{alg:algo_mod1}.

\begin{algorithm}[htbp]
    \caption{Metropolis-Hasting-within-Gibbs sampling scheme for the joint posterior (\ref{eq:aposterioriJointe3_bis})}
    \vspace{0.3cm}
    \label{alg:algo_mod1}
    \begin{algorithmic}[1]
        \State Initialization with $k = 0$ and arbitrary values for $\phi_0$ and $\Sigma_0$

        \Statex
        
        \State  Increment $k = k+1$
        
        \Statex

        \State Draw a sample $\Sigma^{(k)}|\phi^{(k-1)},\beta^{(k-1)},S \sim \mathcal{IW} \left(\beta^{(k-1)}+n,S+\phi^{(k-1)} I \right)$

        \Statex   
        
        \State Draw a sample $\phi^{(k)}|\beta^{(k-1)},\Sigma^{(k)},S \sim \mathcal{G}\left(\frac{p\beta^{(k-1)}}{2},\frac{tr(\Sigma^{(k)-1})}{2}\right)$
        
        \Statex
                
        \State Draw a sample $\beta^{(k)} \sim \pi(\beta^{(k)}|\Sigma^{(k)},\phi^{(k)},S)$, see Algorithm \ref{alg:algo_mod3}

        \Statex
        
        \State Return to 2 except in the case of stop criterion
         \vspace{0.3cm}
    \end{algorithmic}
\end{algorithm}

\noindent Then to obtain samples from the posterior distribution (\ref{eq:aposterioriJointe3_mod2}) of $(\Sigma, \Phi,\beta)$ from Model 2, you need to follow all steps of Algorithm \ref{alg:algo_mod1} except Step 4 and instead, use Step 4$_{bis}$ described in Algorithm \ref{alg:algo_mod2}.

\begin{algorithm}[htbp!]
    \caption{Modification in Algorithm \ref{alg:algo_mod1} for sampling from the joint posterior (\ref{eq:aposterioriJointe3_mod2})}
    \label{alg:algo_mod2}
    \begin{algorithmic}
   
        \State  \hspace*{-3mm} 4$_{bis}$: For $j$ from $1$ to $p$, draw a sample $\phi_j^{(k)}|\beta^{(k-1)},\Sigma^{(k)},S \sim \mathcal{G}\left(\frac{p\beta^{(k-1)}}{2},\frac{\Sigma_{jj}^{(k)-1}}{2}\right)$     
    \end{algorithmic}
\end{algorithm}

\subsection{The Metropolis sampling method for $\b$}
The density of $\b^{(k)}|\Sigma^{(k)},\phi^{(k)},S$ is such that\\
\begin{equation}
\begin{split}
& \pi(\beta^{(k)}|\Sigma^{(k)},\phi^{(k)},S)\ \propto \exp\left[\beta^{(k)} C^{(k)} - \delta \log \beta^{(k)} - \log(\Gamma_p\left(\frac{\beta^{(k)}}{2}\right))\right]\mathbb{I}_{]p+1;+\infty[} (\beta^{(k)})\\
& \text{with } C^{(k)}=\frac{\log \left|\Sigma^{(k)-1}\right|+\log\left|\Phi^{(k)}\right| - p \log 2}{2}.
\label{eq_beta}
\end{split}
\end{equation}
We describe a random-walk Metropolis algorithm to sample from the distribution of $\gamma^{k}$ when $\gamma^{(k)}=log(\b^{(k)}-p-1)$. The proposed algorithm is a Markov chain with a Gaussian random-walk centered in the value $\gamma^{(k-1)}$ as symmetric proposal density $q$ at iteration $k$.
\begin{equation*}
q(\gamma^{prop})=\mathcal{N}_p\left(\gamma^{(k-1)}, 2 \times \hat{\sigma^2}^{(k-1)}\right).
\end{equation*}
$\hat{\sigma^2}^{(k-1)}$ corresponds to the variance of $\gamma^{k}$, estimated by numerical integration. The acceptance probability is equal to $min\left(1,\frac{q(\gamma^{prop})}{q(\gamma^{(k-1)})}\right)$. \\
Samples from (\ref{eq_beta}) can be obtained using the mapping described in Algorithm \ref{alg:algo_mod3}. 

\begin{algorithm}[htbp]
    \caption{The random-walk Metropolis algorithm for the joint posterior (\ref{eq_beta})}
    \vspace{0.3cm}
    \label{alg:algo_mod3}
    \begin{algorithmic}[1]
        \State Set $\gamma^{(k-1)} = \log(\b^{(k-1)} -p-1)$ 
        
        \Statex

        \State Compute $\hat{\sigma^2}^{(k-1)}$ by numerical integration 

        \Statex

        \State Sample $\gamma^{prop}$ from $q(\gamma^{prop} | \gamma^{(k-1)}, 2 \times \hat{\sigma^2}^{(k-1)})$

        \Statex   
        
        \State Set $\log(\rho)= \log q(\gamma^{prop}) - \log q(\gamma^{(k-1)})$
       
        \Statex   
        
        \State Sample $u$ from $\mathcal{U}_{[0;1]}$

        \Statex
                
        \State If $\log \rho> \log u$ then $\gamma^{(k)}=\gamma^{prop}$. Otherwise $\gamma^{(k)}=\gamma^{(k-1)}$
         
        \Statex
                
        \State Set $\b^{(k)} = \exp(\gamma^{(k)})+p+1$
      \vspace{0.3cm}
    \end{algorithmic}
\end{algorithm}

\section{Simulation results}
\label{sec_simulation}
\indent The objective of this section is to give a detailed comparison between different estimators: the proposed Bayes estimators and the maximum likelihood estimator. More specifically, the study will consider the true matrices used in \cite{Daniels99}, with different structures and conditionings\footnote{A matrix is said ill-conditioned if the ratio of its maximum and minimum eigenvalue is large. The closer it is to 1, the better conditioned the matrix is.}. We will derive frequentist characteristics of the Bayesian procedures in terms of risks associated to three loss functions. Through this simulation study, we will get ideas about the behaviour of the competing estimation methods in different situations.

\subsection{Simulation design}
\label{subsec_simu}

We carried out the simulation study from Section 3 of Daniels and Kass's paper \cite{Daniels99}. They consider seven covariance matrices of dimension $p=5$: three diagonal and four non-diagonal matrices. The first, A, is an identity matrix; the second, B, represents a covariance matrix with roughly equally spaced eigenvalues increasing in powers of 0.75 from 0.75$^0$ to 0.75$^4$; the third, C, is a somewhat ill-conditionned  matrix, with eigenvalues equal to  0.75$^0$, 0.75$^1$, 0.75$^2$, 0.75$^{10}$ and 0.75$^{20}$. They are then combined with rotations to produce four full true covariance matrices:	B1 and C1 are matrix B and matrix C with Givens angles all set to $\frac{\pi}{4}$, B2 and C2 are matrixces B and C with Givens angles evenly spaced between (-$\frac{\pi}{4}$,$\frac{\pi}{4}$).\\
\\
From each covariance matrix, we do the following simulation process:
\begin{itemize}
\item simulate a sample of size $n$,
\item compute $\hat{\s}_{1,L_1}$ , $\hat{\s}_{2,L_1}$, $\hat{\s}_{DK,L_1}$, $\hat{\s}_{1,L_2}$, $\hat{\s}_{2,L_2}$ and $\hat{\s}_{DK,L_2}$ respectively the estimators for $\Sigma$ from Model 1 of Section \ref{sec_model_1} (with $\delta=2$), from Model 2 of Section \ref{sec_model_2} (with $\delta=2$), from Model D\&K of Section \ref{sec_model_DK} (with $b=10^6$), under $L_1$ and $L_2$ losses, \item compute $\hat{\s}_{MLE}$,
\item compute the associated loss $L_i(\hat{\s},\s)$ for $i=1,2$ and for each estimator $\hat{\Sigma}$.
\end{itemize}  
These estimations are carried out with the Metropolis-within-Gibbs algorithms described in Section \ref{sec_MCMC} with 20,000 iterations, among which 5,000 are used for the burn-in period. \\
We then compare the different estimators with respect to the risk function $R_i(\hat{\s},\s)=\mathbb{E}_\s \left(L_i(\hat{\s}, \s)\right) \ i=1,2.$
These frequentist risks are approximated by repeating 100 times the previous simulation process.\\
\\
From Propositions 1 and 2, $\delta$ must be strictly greater than $1$. For the remainder of this paper, we will choose, quite arbitrarily, the smallest possible integer for $\delta$, that is, $\delta =2$. As an extension to this work, we could assign a prior on this parameter.

\subsection{Performance comparisons}

Here we proceed to comparison of risks - associated to a specific loss, $L_1$ or $L_2$ - between the prior models. Then we compare the ability to accurately estimate the eigenvalues accross all competing estimators. 

\subsubsection{Under the Frobenius loss}
Table \ref{risks_bis} summarizes the simulation results for the frequentist risk $R_2$ 
for sample sizes $n=5$ and $n=100$. \\
\noindent 
The main remark is that, for $n=5$, the shrinkage estimator from Model 1 always leads to (sometimes dramatic: risk divided by 2) improvement in accuracy over the alternative procedures. Secondly the estimators from Model 2 and Model D\&K fail to accurately estimate and do even worse than the maximum likelihood estimator in all cases. The models in question lead only to shrinkage of the off-diagonal elements and such regularization methods tend to not have the expected beneficial effect of being more precise, under $L_2$ loss. In contrast Model 1 involves more severe shrinkage, which is given a very positive welcome. \\ 
For a bigger sample size, the differences in performance between the estimators disappear except for the identity-matrix case (A).

\begin{table}
\centering
\begin{tabular}{|c|c|c|c|c|} 
\hline
 &  \multicolumn{4}{|c|}{n = 5} \\
\hline
True matrices   &   $\hat{\s}_{1,L_2}$ &    $\hat{\s}_{2,L_2}$ &   $\hat{\s}_{DK,L_2}$ &  $\hat{\s}_{MLE}$ \\
\hline
\multicolumn{ 1}{|c|}{A}& 0.88	(0.13)&	8.29	(0.71)&	7.09	(0.59)&	1.72	(0.12)\\
\multicolumn{ 1}{|c|}{B}& 0.57	(0.06)&	3.87	(0.45)&	3.15	(0.39)&	0.83	(0.08)\\
\multicolumn{ 1}{|c|}{B1}& 0.55	(0.06)&	3.75	(0.33)&	3.22	(0.32)&	0.79	(0.07)\\
\multicolumn{ 1}{|c|}{B2}& 0.39	(0.04)&	3.48	(0.30)&	2.77	(0.23)&	0.73	(0.06)\\
\multicolumn{ 1}{|c|}{C}& 0.57	(0.05)&	2.87	(0.46)&	2.69	(0.78)&	0.62	(0.09)\\
\multicolumn{ 1}{|c|}{C1}& 0.45	(0.05)&	2.39	(0.35)&	2.39	(0.34)&	0.60	(0.09)\\
\multicolumn{ 1}{|c|}{C2}& 0.47	(0.06)&	2.07 (0.27)&	2.17	(0.28)&	0.64	(0.08)\\
\hline
           &  \multicolumn{4}{|c|}{n = 100} \\
\hline
\multicolumn{ 1}{|c|}{A} & 0.03	(0.004)&	0.11	(0.007)&	0.10	(0.007)&	0.09	(0.006)\\
\multicolumn{ 1}{|c|}{B} & 0.04	(0.003)&	0.04	(0.004)&	0.04	(0.004)&	0.04	(0.003)\\
\multicolumn{ 1}{|c|}{B1} & 0.04	(0.003)&	0.04	(0.004)&	0.04	(0.004)&	0.04	(0.004)\\
\multicolumn{ 1}{|c|}{B2} & 0.04	(0.003)&	0.05	(0.004)&	0.04	(0.004)&	0.04	(0.004)\\
\multicolumn{ 1}{|c|}{C} & 0.04	(0.004)&	0.04	(0.004)&	0.04	(0.004)&	0.04	(0.004)\\
\multicolumn{ 1}{|c|}{C1} & 0.03	(0.003)&	0.04	(0.004)&	0.04	(0.004)&	0.03	(0.003)\\
\multicolumn{ 1}{|c|}{C2} & 0.03	(0.002)&	0.03	(0.002)&	0.03	(0.002)&	0.03	(0.002)\\
\hline
\end{tabular}
\caption{\label{risks_bis}Comparison in risk between the Bayes estimators and the usual estimator of covariance matrix under $L_2$. The values in parentheses refer to the simulation standard errors.}
\end{table}

\subsubsection{Under Stein's loss}
Table \ref{risks} gives the simulation results for the frequentist risk $R_1$ 
for sample sizes $n=5$ and $n=100$.\\
When $n$=5, the Bayes estimators provide substantial improvement in risk compared to the sample covariance matrix for the well-conditioned matrices (A, B, B1, B2), with risks from 3.78 to 8.7 times smaller. In these cases, Model 1 does somewhat better than Model 2 and Model D\&K but as soon as $\s$ is ill-conditioned (C, C1, C2), it does worse than all other estimators.  
For the ill-conditioned diagonal matrix (C), the estimators from Model 2 and Model D\&K perform well compared to the sample covariance matrix while they do poorly for its two rotated versions of matrices (C1, C2).   \\ 
As expected, when the sample size becomes large (here $n$=100), the differences in performance between the estimators become blurred. Nevertheless in the well-conditioned cases the Bayes estimators still offer a non-negligible percentage reduction in risk compared to the sample covariance matrix.

\begin{table}
\centering 
\begin{tabular}{|c|c|c|c|c|} 

\hline
&  \multicolumn{4}{|c|}{n = 5} \\
\hline
True matrices            &   $\hat{\s}_{1,L_1}$ &    $\hat{\s}_{2,L_1}$ &   $\hat{\s}_{DK,L_1}$ &  $\hat{\s}_{MLE}$ \\
\hline
\multicolumn{ 1}{|c|}{A}& 0.66	(0.06)&	1.42	(0.07)&	1.18	(0.07)&	5.75	(0.25)\\
\multicolumn{ 1}{|c|}{B}& 0.87	(0.05)&	1.50	(0.09)&	1.26	(0.08)&	5.61	(0.21)\\
\multicolumn{ 1}{|c|}{B1}& 0.77	(0.05)&	1.45	(0.07)&	1.34	(0.07)&	5.77	(0.23)\\
\multicolumn{ 1}{|c|}{B2}& 0.70	(0.04)&	1.47	(0.07)&	1.34	(0.06)&	5.52	(0.21)\\
\multicolumn{ 1}{|c|}{C}&42.42	(0.26)&	1.37	(0.08)&	1.17	(0.07)&	5.69	(0.22)\\
\multicolumn{ 1}{|c|}{C1}& 41.87	(0.26)&	17.16	(0.97)&	26.72	(2.39)&	6.31	(0.32)\\
\multicolumn{ 1}{|c|}{C2}& 42.77	(0.25)&	26.36	(1.97)&	42.04	(3.30)&	6.22	(0.25)\\
\hline
           &  \multicolumn{4}{|c|}{n = 100} \\
\hline
\multicolumn{ 1}{|c|}{A} & 0.03	(0.003)&	0.07	(0.003)&	0.06	(0.003)&	0.15	(0.006)\\
\multicolumn{ 1}{|c|}{B} & 0.11	(0.004)&	0.07	(0.004)&	0.06	(0.004)&	0.15	(0.005)\\
\multicolumn{ 1}{|c|}{B1} & 0.12	(0.005)&	0.13	(0.005)&	0.13	(0.006)&	0.16	(0.006)\\
\multicolumn{ 1}{|c|}{B2} & 0.13	(0.005)&	0.13	(0.005)&	0.14	(0.006)&	0.16	(0.006)\\
\multicolumn{ 1}{|c|}{C} & 0.18	(0.007)&	0.07	(0.004)&	0.06	(0.003)&	0.15	(0.005)\\
\multicolumn{ 1}{|c|}{C1} & 0.18	(0.007)&	0.16	(0.006)&	0.16	(0.006)&	0.16	(0.006)\\
\multicolumn{ 1}{|c|}{C2} & 0.19	(0.006)&	0.17	(0.006)&	0.16	(0.006)&	0.16	(0.006)\\
\hline
\end{tabular}  
\caption{\label{risks}Comparison between estimators with respect to the risk function $R_1$. The values in parentheses refer to the simulation standard errors.}
\end{table}

\subsubsection{About eigenvalues estimation}
An important issue in covariance matrix estimation is the bias of the estimators of the extreme eigenvalues. We can then compare the bias in the eigenvalues of estimates based on the different models and approaches and we try to enlighten the previous results.\\ 
We set  
\begin{equation}
L_{\lambda_i}(\lambda_i,\hat{\lambda}_i)=\frac{\left|\lambda_i - \hat{\lambda}_i \right| }{\lambda_i}, \text{   with } i \in \left\{min,max\right\}
\end{equation}
as a relative measure of distance between the true ($\lambda_i$) and inferred eigenvalue ($\hat{\lambda}_i$).\\
Tables \ref{vpMax_comp} and \ref{vpMin_comp} give results for the minimal ($\lambda_{min}$) and maximal ($\lambda_{max}$) eigenvalues of the seven estimators for each type of covariance matrix. The eigenvalues of $\hat{\s}_{MLE}$ differ greatly from the true values, especially when $\s$ is close to the identity matrix. $\hat{\s}_{1,L_1}$, $\hat{\s}_{2,L_1}$, $\hat{\s}_{DK,L_1}$ and $\hat{\s}_{1,L_2}$ appear to successfully estimate $\lambda_{max}$ in all cases but fail in estimating $\lambda_{min}$ in case of a strongly misspecified target. $\hat{\s}_{2,L_2}$ and $\hat{\s}_{DK,L_2}$ can improve $\lambda_{min}$ but is always inappropriate for an accurate estimation of $\lambda_{max}$.\\
These results, combined with the boxplots of the smallest and largest eigenvalues, presented Annexe \ref{annexe_eigenvalue_comp}, highlight that the Bayes methods reduce the distorsion of the eigenvalue spectrum but that this effect can be too pronounced. Moreover the results stress that the performances under $L_1$ loss can be explained by the eigenvalues shrinkage: overskhrinkage of the small eigenvalues will imply a poor performance under $L_1$ loss whereas it will not be penalized by $L_2$ loss which focuses on the errors on big values.\\
Estimator derived from Model 1 will be forced to be well-conditioned. Consequently it will yield to overshrinkage as soon as the true matrix has its eigenvalues far apart, but only of the smallest eigenvalues. \\
The overestimation phenomenom can be explained by the lower bound of the hyperparameter $\beta$ as illustrated in Section \ref{sec_etude_beta}.

\begin{table}
\centering
\begin{tabular}{|r|c|c|c|c|c|c|r|}
\hline
           &   $\hat{\s}_{1,L_1}$ &    $\hat{\s}_{2,L_1}$ &   $\hat{\s}_{DK,L_1}$ &    $\hat{\s}_{1,L_2}$ &    $\hat{\s}_{2,L_2}$ &   $\hat{\s}_{DK,L_2}$ &  $\hat{\s}_{MLE}$ \\
\hline
         A &       0.48 (0.02)&       0.73 (0.02)&       0.67 (0.02)&  0.25 (0.02)&       0.45 (0.02)&       0.46 (0.02)&       0.97 (0.01)\\
\hline
         B & 0.35 (0.03)&       0.58 (0.02)&       0.50 (0.03)&       0.65 (0.06)&       0.43 (0.03)&       0.44 (0.03)&       0.95 (0.01)\\
\hline
        B1 &  0.29 (0.02)&       0.52 (0.02)&       0.45 (0.02)&       0.71 (0.05)&       0.50 (0.04)&       0.49 (0.04)&       0.95 (0.01)\\
\hline
        B2 &  0.29 (0.02)&       0.54 (0.02)&       0.47 (0.02)&       0.67 (0.05)&       0.45 (0.04)&       0.42 (0.03)&       0.95 (0.01)\\
\hline
         C &      43.58 (2.62)&  0.43 (0.03)& 0.43 (0.04)&      93.67 (4.47)&       1.01 (0.09)&       0.98 (0.10)&       0.79 (0.02)\\
\hline
        C1 &      43.15 (2.63)&      11.47 (0.74)&      13.96 (1.05)&      94.28 (4.27)&      36.87 (2.12)&      40.38 (2.45)&  0.86 (0.02)\\
\hline
        C2 &      44.06 (2.50)&      19.26 (1.10)&      26.82 (1.89)&      97.61 (4.64)&      65.21 (3.48)&      76.81 (4.59)& 0.85 (0.02)\\
\hline
\end{tabular}  
\caption{\label{vpMin_comp}Comparison in risk under $L_{\lambda_{min}}$, between the smallest eigenvalue of the competing estimators when $n=5$. The values in parentheses refer to the standard errors.}
\end{table}

\begin{table}
\centering
\begin{tabular}{|r|c|c|c|c|c|c|r|}
\hline
           &    $\hat{\s}_{1,L_1}$ &    $\hat{\s}_{2,L_1}$ &   $\hat{\s}_{DK,L_1}$ &    $\hat{\s}_{1,L_2}$ &    $\hat{\s}_{2,L_2}$ &   $\hat{\s}_{DK,L_2}$ &  $\hat{\s}_{MLE}$ \\
\hline
\multicolumn{ 1}{|c|}{A} &  0.27 (0.02) &       0.59 (0.04) &       0.64 (0.05)&       0.84 (05)&       2.26 (0.10)&       1.99 (0.10)&       1.70 (0.08) \\
\hline
\multicolumn{ 1}{|c|}{B} &       0.34 (0.02)&       0.37 (0.03)&       0.40 (0.06) &  0.33 (0.03)&       1.21 (0.10) &       1.02 (0.09) &       0.68 (0.06) \\
\hline
\multicolumn{ 1}{|c|}{B1} &       0.29 (0.02)& 0.28 (0.02)&       0.31 (0.03)&       0.37 (0.04)&       1.24 (0.08)&       1.03 (0.08)&       0.84 (0.07)\\
\hline
\multicolumn{ 1}{|c|}{B2}  &       0.29 (0.02)&  0.27 (0.02)&       0.29 (0.02)&       0.30 (0.02)&       1.14 (0.07)&       0.94 (0.06)&       0.71 (0.05)\\
\hline
\multicolumn{ 1}{|c|}{C} &       0.41 (0.02)&  0.33 (0.03)&       0.34 (0.04) &       0.38 (0.03)&       1.10 (0.09)&       0.96 (0.10)&       0.57 (0.05)\\
\hline
\multicolumn{ 1}{|c|}{C1}  &       0.38(0.02) &  0.31 (0.02)&       0.33 (0.03)&       0.42 (0.04)&       1.06 (0.08)&       1.02 (0.09)&       0.66 (0.06)\\
\hline
\multicolumn{ 1}{|c|}{C2}  &       0.37 (0.02)&  0.31 (0.02)&       0.34 (0.02)&       0.40 (0.04)&       0.93 (0.07)&       0.90 (0.08)&       0.67 (0.06)\\
\hline
\end{tabular}  
\caption{\label{vpMax_comp}Comparison in risk under $L_{\lambda_{max}}$, between the largest eigenvalue of the competing estimators when $n=5$.  
The values in parentheses refer to the simulation standard errors.}
\end{table}

\subsubsection{Posterior distribution of $\b$}
\label{sec_etude_beta}
\begin{figure}[!]
	\centering
		\includegraphics[width=0.7\textwidth]{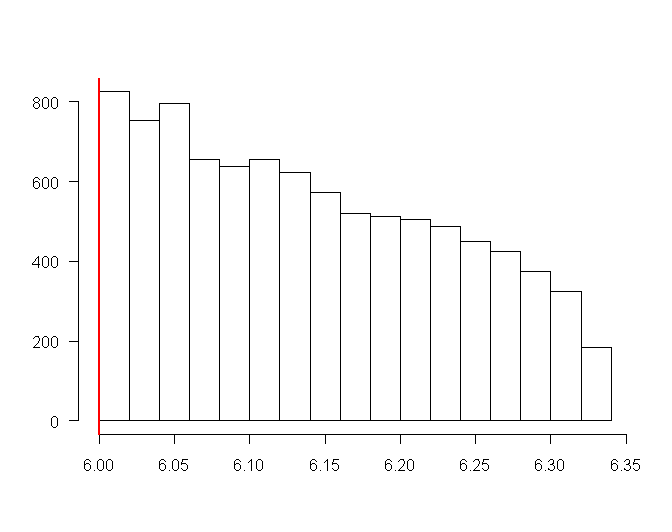}	
	\caption{\label{fig:beta_posterior_n100}Histogram of the posterior samples of $\beta$ from (\ref{eq:aposterioriJointe3_mod2}) when $n$=100 in the full ill-conditioned case (C2). The red line represents the lower bound of $\b$. 
}
\end{figure}

As seen in Section \ref{sec_model}, the hyperparameter $\beta$ is known to control the amount of shrinkage. As the use of hierarchical models allows to estimate $\beta$ from data, data will determine the shrinkage intensity. Therefore, if the structured scale matrix is close to the true matrix, then $\beta$ would take high values. Inversely one expects to get values for $\beta$ arbitrarily small in the case where data support structure far from the prior scale matrix configuration. However, as mentioned by Daniel and Kass \cite{Daniels99}, a low intensity for shrinkage will be impossible to obtain 
because $\beta$ must be always bounded by $p-1$ minimum.\\
For illustration Figure \ref{fig:beta_posterior_n100} represents the histogram of posterior samples of $\beta$ issued from Model 2 (a sample path of length 15,000 for one of the 100 datasets) when the true covariance matrix is C2 and $n=100$. The posterior distribution of $\b$ is very concentrated on the lower bound $p$+1. 

\section{Conclusion and discussion}

\indent In this paper we proposed hierarchical Bayesian shrinkage methods for the estimation of covariance matrices in a small sample setting. Inverse-Wishart priors were considered with unknown hyperparameters. We focused on a "diagonal, common variance" and two "diagonal, unequal variance" models as covariance targets. Models are with 2 to $p$+1 free parameters, on which we assigned noninformative priors. We showed in details the conditions to ensure the property of the posterior distributions and proposed a Metropolis-Hasting-within-Gibbs algorithm to sample from them. Then we gave a detailed comparison between the different Bayesian estimators and the classical maximum likelihood covariance estimator, under three loss functions.\\
As statistically efficient and computationally fast alternative to the widely used standard covariance estimators, we recommend the shrinkage covariance estimators which shrink all components of the empirical covariance matrix, that is not only perfectly applicable to small samples but can also improve the classical estimators for large $n$. These improved estimators exhibit none of the defects of the standard covariance estimators, in particular they reduce variance, they are always positive definite and well-conditioned. This property might imply overestimation of the small eigenvalue. By producing a well-conditioned positive definite covariance estimate one automatically also obtains an equally well-conditioned estimate of the inverse covariance - a quantity of crucial importance, for instance, in classification or graphical models. For other goals like reduction dimension, principal component analysis needs to successfully estimate the largest eigenvalues rather than the smallest ones. However, for applications where we have to focus on estimation of all eigenvalues, care must be taken. Indeed under Stein's loss function, the proposed models can be unefficient. This happens when the data is in conflict with the specified prior structure, it is directly due to the lower bound of the degrees of freedom of an inverse-Wishart prior. Overshrinkage of the small eigenvalues is then caused.\\
The evidence on differences in estimation performance of different estimation approaches and models suggest that there is no "best approach" and that the relative accuracy of one approach or model in comparison to another depends strongly on the problem, that's one of the basic principles that underlie the Bayes paradigm. \\
A direct perspective of this work would be to investigate more complex models to overcome the issue of overshrinkage. 
Further work will be to investigate the effects of the estimation procedures in real data on the Value at Risk (VaR) computation. The VaR is a widely used tool for risk assessment in finance and is defined as a quantile of the predictive probability distribution for the amount of a future financial loss. 
The standard method for approximating the VaR is based on calculations using Monte-Carlo simulations of asset prices from a Gaussian distribution with unknown covariance matrix. The classical estimators have no full rank. Besides, in this application, focus is on the largest eigenvalues and then the proposed shrinkage estimator seems to be appropriate.

\section*{Acknowledgments}

We wish to thank Christian Robert and Jean-Michel Marin for their contribution. \\
We would like to thank EDF (company of Electricity of France), especially Financial Direction, to support the thesis of Mathilde Bouriga.

\newpage

\bibliographystyle{elsarticle-harv}
\bibliography{main}

\begin{thebibliography}{28}
\expandafter\ifx\csname natexlab\endcsname\relax\def\natexlab#1{#1}\fi
\expandafter\ifx\csname url\endcsname\relax
  \def\url#1{\texttt{#1}}\fi
\expandafter\ifx\csname urlprefix\endcsname\relax\def\urlprefix{URL }\fi

\bibitem[{Barnard et~al.(2000)Barnard, McCulloch, and Meng}]{Bar00}
Barnard, J., McCulloch, R., Meng, X., 2000. Modeling covariance matrices in
  terms of standard deviations and correlations, with applications to
  shrinkage. Statistica Sinica 10, 1281--1311.

\bibitem[{Champion(2001)}]{Champion01}
Champion, C., 2001. {Empirical Bayesian estimation of normal variances and
  covariances}. Journal of Multivariate Analysis 26~(2), 60--79.

\bibitem[{Daniels and Kass(1999)}]{Daniels99}
Daniels, M., Kass, R., 1999. Nonconjugate bayesian estimation of covariance
  matrices and its use in hierarchical models. Journal of the American
  Statistical Association 94~(448), 1254--1263.

\bibitem[{Dempster(1969)}]{Dem69}
Dempster, A., 1969. Elements of Continuous Multivariate Analysis.
  Addison-Wesley, Reading, Mass.

\bibitem[{Dey and Srinivasan(1985)}]{Dey85}
Dey, D., Srinivasan, C., 1985. Estimation of a covariance matrix under
  {S}tein's loss. The Annals of Statistics 13~(4), 1581--1591.

\bibitem[{Diaconis and Ylvisaker(1979)}]{Diac79}
Diaconis, P., Ylvisaker, D., 1979. Conjugate priors for exponential families.
  The Annals of Statistics 7~(2), 269--281.

\bibitem[{Eaton(1983)}]{Eat83}
Eaton, M., 1983. Multivariate Statistics : A Vector Space Approach. Wiley.

\bibitem[{Efron and Morris(1974)}]{Efron74}
Efron, B., Morris, C., 1974. Multivariate empirical bayes and estimation of
  covariance matrices. The Annals of Statistics 4, 22--32.

\bibitem[{Franck and Zakoïan(2000)}]{Fran00}
Franck, C., Zakoïan, J., 2000. Covariance matrix estimation for estimators of
  mixing weak arma models. Journal of statistical planning and inference
  83~(2), 369--394.

\bibitem[{Furrer and Bengtsson(2007)}]{Furrer07}
Furrer, R., Bengtsson, T., 2007. Estimation of high-dimensional prior and
  posteriori covariance matrices in kalman filter variants. Journal of
  Multivariate Analysis 98, 227--255.

\bibitem[{Gelman(2006)}]{Gelman06}
Gelman, A., 2006. Prior distributions for variance parameters in hierarchical
  models. Bayesian Analysis 1~(3), 515--533.

\bibitem[{Haff(1979)}]{Haff79}
Haff, L., 1979. Estimation of the inverse covariance matrix : random mixtures
  of the inverse wishart matrix and the identity. The Annals of Statistics
  7~(6), 1264--1276.

\bibitem[{Haff(1980)}]{Haff80}
Haff, L., 1980. Empirical bayes estimation of the multivariate normal
  covariance matrix. The Annals of Statistics 8~(3), 586--597.

\bibitem[{James and Stein(1961)}]{James61}
James, W., Stein, C., 1961. Estimation with quadratic loss. Proceedings of the
  Fourth Berkeley Symposium on Mathematical and Statistical Probabilities Vol.
  1, University of California Press, Berkeley, 361--379.

\bibitem[{Jeffreys(1961)}]{Jef61}
Jeffreys, H., 1961. Theory of probability. 3rd ed. Oxford Classic Texts in the
  Physical Sciences, Oxford: Oxford University Press.

\bibitem[{Kubokawa(2004)}]{Kub04}
Kubokawa, T., 2004. A revisit to estimation of the precision matrix of the
  {W}ishart distribution. Unpublished discussion paper.

\bibitem[{Kubokawa and Srivastava(1999)}]{Kub99}
Kubokawa, T., Srivastava, M., 1999. Estimating the covariance matrix : a new
  approach. Unpublished discussion paper.

\bibitem[{Ledoit and Wolf(2002)}]{Ledoit02}
Ledoit, O., Wolf, M., 2002. Improved estimation of the covariance matrix of
  stock returns with an application to protfolio selection. Journal of
  empirical finance 10, 603--621.

\bibitem[{Leonard and Hsu(1992)}]{Hsu92}
Leonard, T., Hsu, J., 1992. Bayesian inference for a covariance matrix. The
  Annals of Statistics 20~(4), 1669--1696.

\bibitem[{Lin and Perlman(1984)}]{Lin84}
Lin, S., Perlman, M., 1984. A monte carlo comparison of four estimators of
  covariance matrix. Tech. Rep.~44, University of Washington, Dept.of
  statistics.

\bibitem[{Robert(2001)}]{Rob05}
Robert, C., 2001. The Bayesian Choice: from Decision-Theoretic Motivations to
  Computational Implementation. Springer.

\bibitem[{Robert and Marin(2007)}]{Mar07}
Robert, C., Marin, J., 2007. Bayesian core. Springer-Verlag, New York.

\bibitem[{Roberts and Rosenthal(2006)}]{Rob06}
Roberts, G., Rosenthal, J., 2006. Harris recurrence of metropolis-within-gibbs
  and trans-dimensional markov chains. The Annals of Applied Statistics 16~(4),
  2123--2139.

\bibitem[{Schäfer and Strimmer(2005)}]{Scha05}
Schäfer, J., Strimmer, K., 2005. A shrinkage approach to large-scale covariance
  matrix estimation and implications for functional genomics. Statistical
  applications in genetic and molecular biology 4~(32).

\bibitem[{Smith and Kohn(2002)}]{Kohn02}
Smith, M., Kohn, R., 2002. Parsimonious covariance matrix estimation for
  longitudinal data. Journal of the American Statistical Association 97~(460),
  1140--1153.

\bibitem[{Stein(1956)}]{Stein56}
Stein, C., 1956. Some problems in multivariate analysis. Tech. Rep.~6,
  Standford University, Dept. of Statistics.

\bibitem[{Stein(1975)}]{Stein75}
Stein, C., 1975. Estimation of a covariance matrix. Rietz lecture, 39th annual
  meeting IMS. Atlanta, Georgia.

\bibitem[{Yang and Berger(1994)}]{Berger94}
Yang, R., Berger, J., 1994. Estimation of a covariance matrix using the
  reference prior. The Annals of Statistics 22~(3), 1195--1211.

\end{thebibliography}
    
\newpage

\appendix

\section{Proof of Proposition 2}
\label{annexe_loi_propre}
We want to evaluate whether 
\begin{eqnarray} 
\int^{+\infty}_{p+1} \int_{]0;+\infty[^p} \int_{\textsl{S}^+} \pi(\Sigma,\Phi,\beta  | S) d\Sigma d\phi_1...d\phi_p  d\beta 
\label{eq:cste_post_joint}
\end{eqnarray} 
is finite or not. \\
By Fubini's theorem:
\begin{eqnarray} 
\int^{+\infty}_{p+1} \int_{]0;+\infty[^p} \int_{\textsl{S}^+} \pi(\Sigma,\Phi,\beta  | S) d\Sigma d\phi_1...d\phi_p d\beta  
= \int^{+\infty}_{p+1}\pi(\beta  | S) d\beta.
\label{eq:fubini_application}
\end{eqnarray}  

\noindent Thus it is sufficient to prove the convergence of the posterior marginal distribution of $\beta$.\\
\\
We begin to marginalize the density $\pi(\Sigma,\Phi,\beta  | S)$ over $\Sigma$ to obtain an expression for $\pi(\Phi, \beta | S)$. We have: 

\begin{equation*}
\pi(\Phi,\beta | S)  \propto \frac{\Gamma_p(\frac{\beta+n}{2})}{\Gamma_p(\frac{\beta}{2})} \frac{|\Phi|^{\frac{\beta}{2}-1}}{|S+\Phi|^{\frac{\beta+n}{2}}}\frac{1}{\beta^{\delta}}.
\end{equation*}
Denote $Q\Lambda Q^T$, the spectral decomposition of the matrix $S$. Then 
\[ \begin{array}{rcl}
\pi(\Phi,\beta | S)  & \propto & \frac{\Gamma_p(\frac{\beta+n}{2})}{\Gamma_p(\frac{\beta}{2})} \frac{|\Phi|^{\frac{\beta}{2}-1}}{|Q\Lambda Q^T+\Phi|^{\frac{\beta+n}{2}}}\frac{1}{ \beta^{\delta}}\\
\\
& \propto & \frac{\Gamma_p(\frac{\beta+n}{2})}{\Gamma_p(\frac{\beta}{2})} \frac{|\Phi|^{\frac{\beta}{2}-1}}{|Q\Lambda Q^T+ Q \Phi Q^T|^{\frac{\beta+n}{2}}}\frac{1}{\beta^{\delta}}\\
\\
& \propto & \frac{\Gamma_p(\frac{\beta+n}{2})}{\Gamma_p(\frac{\beta}{2})} \frac{|\Phi|^{\frac{\beta}{2}-1}}{|Q (\Lambda+\Phi) Q^T|^{\frac{\beta+n}{2}}}\frac{1}{\beta^{\delta}}\\
\\
& \propto & \frac{\Gamma_p(\frac{\beta+n}{2})}{\Gamma_p(\frac{\beta}{2})} \frac{|\Phi|^{\frac{\beta}{2}-1}}{(|QQ^T| |\Lambda+\Phi|)^{\frac{\beta+n}{2}}}\frac{1}{ \beta^{\delta}}\\
\\
& \propto & \frac{\Gamma_p(\frac{\beta+n}{2})}{\Gamma_p(\frac{\beta}{2})} \frac{\prod^{p}_{i=1} \phi_i^{\frac{\beta}{2}-1}}{\prod^{p}_{i=1}(\lambda_i+\phi_i )^{\frac{\beta+n}{2}}}\frac{1}{\beta^{\delta}}.
\end{array} \] 
Hence 
$\pi(\beta | S) \propto  \int_{]0;+\infty[^p} \frac{\Gamma_p(\frac{\beta+n}{2})}{\Gamma_p(\frac{\beta}{2})} \frac{\prod^{p}_{i=1} \phi_i^{\frac{\beta}{2}-1}}{\prod^{p}_{i=1}(\lambda_i+\phi_i )^{\frac{\beta+n}{2}}}\frac{1}{\beta^{\delta}} \ \prod^{p}_{i=1} d\phi_i$. \\
\\We just need to find a density $g(\beta)$ which dominates the positive function $\pi(\beta | S)$ on $] p+1 ; +\infty [$ and define the conditions so that the latter converges.\\
Let $\lambda_{min}$ be the smallest eigenvalue of $S$, it follows: 
\begin{eqnarray}
\pi(\beta | S) \ & \leq & \ \int_{]0;+\infty[^p} \frac{\Gamma_p(\frac{\beta+n}{2})}{\Gamma_p(\frac{\beta}{2})} \frac{\prod^{p}_{i=1} \phi_i^{\frac{\beta}{2}-1}}{\prod^{p}_{i=1}(\lambda_{min}+\phi_i )^{\frac{\beta+n}{2}}}\frac{1}{\beta^{\delta}} \prod^{p}_{i=1} d\phi_i \ \ \forall \ \beta \ \in \ ] p+1 ; +\infty [ \nonumber \\
\nonumber \\
 & = & \ \pi(\beta | S=\lambda_{min}I). 
 \label{eq:inegalite}
\end{eqnarray}

\vspace{0.3cm}
\noindent Let us check if $\pi(\beta | S=\lambda_{min}I)$ is an integrable function over $]p+1;+\infty[$.\\
\\
We have 
\[ \begin{array}{rcl}
\int_{]0;+\infty[^p} \pi(\phi_1,...,\phi_p,\beta | S=\lambda_{min}I)  \prod^{p}_{i=1} d\phi_i & \propto &  
\int_{]0;+\infty[^p} \frac{\Gamma_p(\frac{\beta+n}{2})}{\Gamma_p(\frac{\beta}{2})} \sqrt{\frac{ \prod^{p}_{i=1}\phi_i^{\beta}}{\prod^{p}_{i=1}(\phi_i+\lambda_{min})^{\beta+n}}}\frac{1}{\prod^{p}_{i=1}\phi_i}\frac{1}{\beta^{\delta}} \prod^{p}_{i=1} d\phi_i\\ 
\\
& \propto & 
\frac{\Gamma_p(\frac{\beta+n}{2})}{\Gamma_p(\frac{\beta}{2})} \frac{1}{\beta^{\delta}} \ 
\prod^{p}_{i=1} \int^{+\infty}_{0}  \sqrt{\frac{\phi_i^{\beta}}{\left[\frac{\lambda_{min}}{n}( \frac{n}{\lambda_{min}}\phi_i+n)\right]^{\beta+n}}}\frac{1}{\phi_i} d\phi_i  \\
\\
& \propto & 
\frac{\Gamma_p(\frac{\beta+n}{2})}{\Gamma_p(\frac{\beta}{2})} \frac{1}{\beta^{\delta}} \ 
\prod^{p}_{i=1} \int^{+\infty}_{0}  \sqrt{\frac{\phi_i^{\beta}}{\left[ \frac{\lambda_{min}}{n}(\beta\frac{n}{\lambda_{min} \beta}\phi_i+n)\right]^{\beta+n}}}\frac{1}{\phi_i} d\phi_i 
\label{eq:aposteriori_beta}
\end{array} \] 

\vspace{0.3cm}

\noindent In each single integral over $\phi_i$, the density of a Fisher F-distribution with $\beta$ and $n$ degrees of freedom appears for the random variables $\frac{n \phi}{\lambda_{min} \beta}$. It implies:

\[ \begin{array}{rrll}

& & \int^{+\infty}_{0} \sqrt{\frac{\left(\frac{n\phi_i}{\lambda_{min}}\right)^{\beta} n^{n}}{\left(\frac{n\phi_i}{\lambda_{min}} + n\right)^{n+\beta}}}\frac{ \frac{n}{\beta \lambda_{min}}}{\frac{n}{\beta\lambda_{min}}\phi_i Beta\left(\frac{\beta}{2},\frac{n}{2}\right)} d\phi_i \\
\\
&  = & \ \int^{+\infty}_{0} \sqrt{\frac{\phi_i^{\beta}\lambda_{min}^{n}}{\left(\phi+ \lambda_{min}\right)^{n+\beta}}}\frac{1}{\phi Beta\left(\frac{\beta}{2},\frac{n}{2}\right)}d\phi_i & = \ 1 
\end{array} \] 

\noindent and thus leads to
\begin{eqnarray}
\pi(\beta | S=\lambda_{min}I) & \propto & \frac{\Gamma_p\left(\frac{\beta+n}{2}\right)}{\Gamma_p\left(\frac{\beta}{2}\right)} \prod^{p}_{i=1} Beta\left(\frac{\beta}{2},\frac{n}{2}\right)\frac{1}{\beta^{\delta}}\nonumber\\
& \propto & \frac{\Gamma_p\left(\frac{\beta+n}{2}\right)}{\Gamma_p\left(\frac{\beta}{2}\right)}\frac{\Gamma(\frac{\beta}{2})^p\Gamma\left(\frac{n}{2}\right)^p}{\Gamma\left(\frac{\beta+n}{2}\right)^p}\frac{1}{\beta^{\delta}}. \nonumber
\label{eq:post_marg_beta}
\end{eqnarray} 

\vspace{0.3cm}

\noindent From the integration properties for positive functions, the inequality (\label{eq:inegalite}) implies the same inequality for their integrals. Hence
\begin{eqnarray}
\int^{+\infty}_{p+1} \pi(\beta | S) \ \leq \ \int^{+\infty}_{p+1} \int_{]0;+\infty[^p} \pi(\phi_1, ..., \phi_p,\beta | S=\lambda_{min}I) \prod^{p}_{i=1} d\phi_i d\beta. \nonumber
\label{eq:inegalite2}
\end{eqnarray}
This function is well-defined in $p+1$. Let us see the behaviour in $+ \infty$.\\

\noindent By definition $\Gamma_p(\beta)  =  \pi^{\frac{p(p-1)}{4}}\prod^{p}_{j=1}\Gamma(\beta+\frac{1-j}{2})$. Furthermore Stirling's formula provides an approximation for the Gamma function: $\Gamma(\beta) \underset{\beta \rightarrow +\infty}{\sim}   \exp^{-\beta}\beta^{\beta-1/2}(2\pi)^{1/2}$. In consequence we get
\[ \begin{array}{rcl}
\Gamma_p(\beta) & \underset{\beta \rightarrow +\infty}{\sim}  & \pi^{\frac{p(p-1)+2p}{4}}2^{p/2}\exp\left(\sum^{p}_{j=1}\frac{j-1}{2}\right)\exp\left(-p\beta\right)\prod^{p}_{j=1}(\beta+\frac{1-j}{2})^{\beta-\frac{j}{2}}.
\end{array} \] 
Hence 
\begin{eqnarray*}
\frac{\Gamma_p(\frac{\beta+n}{2})}{\Gamma_p(\frac{\beta}{2})} & =& \frac{\prod^{p}_{j=1}\Gamma(\frac{\beta+n+1-j}{2})}{ \prod^{p}_{j=1}\Gamma(\frac{\beta+1-j}{2})}\\
& \underset{\beta \rightarrow +\infty}{\sim} & \frac{\exp(-p\frac{\beta+n}{2})\prod^{p}_{j=1}(\frac{\beta+n-1+j}{2})^{\frac{\beta+n-j}{2}}}
{\exp(-p\frac{\beta}{2})\prod^{p}_{j=1}(\frac{\beta-1+j}{2})^{\frac{\beta-j}{2}}}\\
& \underset{\beta \rightarrow +\infty}{\sim} & \frac{\prod^{p}_{j=1}\beta^{\frac{\beta+n-j}{2}}}{2^{pn/2}\exp(\frac{pn}{2})\prod^{p}_{j=1}\beta^{\frac{\beta-j}{2}}}\\ 
& \underset{\beta \rightarrow +\infty}{\sim} &  \frac{\beta^{\frac{pn}{2}}}{2^{pn/2}\exp(\frac{pn}{2})}.
\end{eqnarray*} 

\noindent Moreover 

\begin{eqnarray*}
\frac{\Gamma(\frac{\beta}{2})}{\Gamma(\frac{\beta+n}{2})} & \underset{\beta \rightarrow +\infty}{\sim} & \frac{\exp(-\frac{\beta}{2})(\frac{\beta}{2})^{\frac{\beta-1}{2}}}{\exp(-\frac{(\beta+n)}{2})\left(\frac{(\beta+n)}{2}\right)^{\frac{(\beta+n)-1}{2}}}\\
& \underset{\beta \rightarrow +\infty}{\sim} & \frac{\exp(\frac{n}{2})}{\left(\frac{\beta}{2}\right)^{\frac{n}{2}}}.
\end{eqnarray*} 

\noindent As a result
\begin{eqnarray*}
\pi(\beta | S=\lambda_{min}I) & \propto & \frac{\Gamma_p(\frac{\beta+n}{2})}{\Gamma_p(\frac{\beta}{2})}\left(\frac{\Gamma(\frac{\beta}{2})}{\Gamma(\frac{\beta+n}{2})} \right)^p\frac{1}{\beta^{\delta}} \\
&  \underset{\beta \rightarrow +\infty}{\sim} &  \frac{\beta^{\frac{pn}{2}}}{2^{pn/2}\exp(\frac{pn}{2})}{\frac{\exp(\frac{pn}{2})}{(\frac{p}{2}\beta)^{\frac{pn}{2}}}}\frac{1}{\beta^{\delta}}\\
& \underset{\beta \rightarrow +\infty}{\sim} & \frac{1}{\beta^{\delta}}. 
\end{eqnarray*} 

\noindent Thus, by (\ref{eq:inegalite2}), $\pi(\beta | S)$ is integrable as soon as $\delta > 1$. Together with (\ref{eq:fubini_application}), the same conclusion holds for $\pi(\Sigma,\phi,\beta  | S)$ so that (\ref{eq:cste_post_joint}) is finite.

\section{Complement for simulation results}
\label{annexe_eigenvalue_comp}
These boxplots show the shrinkage effects on the estimates for the extreme eigenvalues.

\begin{figure}[!]
	\centering
		\includegraphics[width=1.1\textwidth]{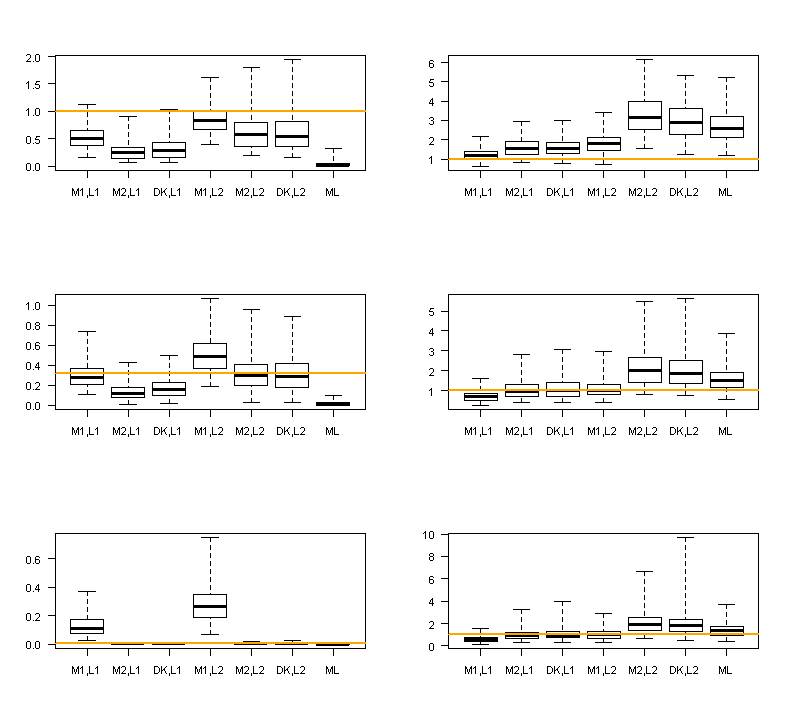}	
	\caption{\label{fig:eigenvalues_L1}Boxplots of the smallest (left) and largest (right) eigenvalues estimates when the true matrices are diagonal. 
The horizontal lines represent the true values of the eigenvalues. 
Top-to-bottom: A, the identity matrix - B, the well-conditioned diagonal matrix - C, the ill-conditioned diagonal matrix.}
\end{figure}

\begin{figure}[!]
	\centering
		\includegraphics[width=1.1\textwidth]{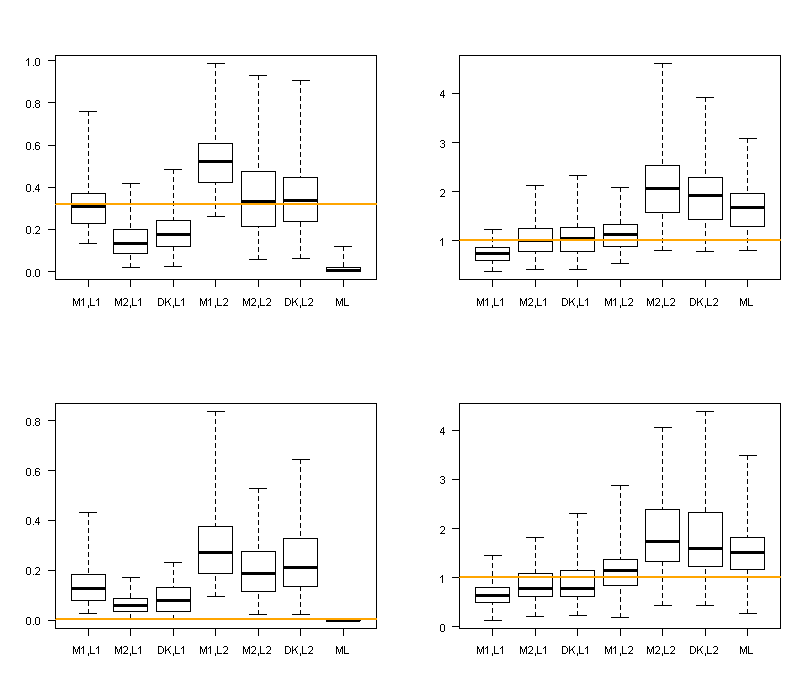}
	 \caption{\label{fig:eigenvaluesuite_L1}Boxplots of the smallest (left) and largest (right) eigenvalues estimates when the true matrices are full. The horizontal lines represent the true values of the eigenvalues. 
Top-to-bottom: B2, the well-conditioned full matrix - C2, the ill-conditioned full matrix.}
\end{figure}

\end{document}